%% file: main.tex
\documentclass[runningheads]{llncs}

\usepackage{url}
\usepackage{svg}
\usepackage{amsmath,amsfonts}
\usepackage{amssymb}

\usepackage{listings}
\usepackage{color}
\usepackage{xcolor}
\definecolor{dkgreen}{rgb}{0,0.6,0}
\definecolor{gray}{rgb}{0.5,0.5,0.5}
\definecolor{mauve}{rgb}{0.58,0,0.82}
\usepackage[ruled,linesnumbered]{algorithm2e}
\usepackage{tabularx}
\usepackage{multirow}
\usepackage{booktabs}
\usepackage{float}
\usepackage{algorithmic}
\algsetup{linenosize=\scriptsize}
\usepackage{setspace}
\usepackage{array}
\usepackage{textcomp}
\usepackage{stfloats}
\usepackage[most]{tcolorbox}
\usepackage{verbatim}
\usepackage{tabularx}
\usepackage{makecell}
\usepackage{bbding}
\usepackage{graphicx}
\usepackage{balance}
\usepackage{threeparttable}
\usepackage{caption}
\usepackage[final]{pdfpages}
\usepackage[commandnameprefix=always, defaultcolor=red]{changes} 
\usepackage[labelformat=simple]{subcaption}

\usepackage{booktabs,multirow,bm,wrapfig,mathtools,dashbox,pifont,soul,enumitem}
\usepackage[commandnameprefix=always, defaultcolor=red]{changes} 

\RequirePackage{mdframed}
\renewmdenv[leftmargin=0.4em, rightmargin=0.4em%
innerleftmargin=0.6em,innerrightmargin=0.6em]{quote}

\newcommand{\AUGP}[1]{AUGP}
\newcommand{\APDETECT}[1]{APDetect}

\setlist[itemize]{leftmargin=*}
\setlist[enumerate]{leftmargin=*}
\newlist{steps}{enumerate}{1}
\setlist[steps, 1]{label = \textbf{RQ\arabic*.}}

\usepackage{hyperref}
\usepackage{url}
\usepackage{xcolor,colortbl,color}
\hypersetup{
 colorlinks=true,
 linkcolor=purple,
 citecolor=violet,
 filecolor=magenta,      
 urlcolor=cyan,
}

\setlist[itemize]{leftmargin=*}
\setlist[enumerate]{leftmargin=*}

\begin{document}
\title{Towards Practical Requirement Analysis and Verification: A Case Study on Software IP Components in Aerospace Embedded Systems}

\author{Zhi Ma\inst{1,2} \and Cheng Wen\inst{2} \and Jie Su\inst{2} \and Ming Zhao\inst{3} \and Bin Yu\inst{1,{\star}} \and Xu Lu\inst{1} \and Cong Tian\inst{1,}\thanks{Corresponding authors: Cong Tian and Bin Yu}}
\institute{ICTT and ISN Laboratory, Xidian University, China \and Guangzhou Institute of Technology, Xidian University, China \and Beijing Institute of Control Engineering, China \\
}

\maketitle

\vspace{-25pt}
\begin{abstract}
\input{Tex/0-abstract}
\end{abstract}

\vspace{-25pt}
\section{Introduction}\label{sec:intro}
\vspace{-5pt}
\input{Tex/1-introduction}

\vspace{-7.5pt}
\section{Background and IP Example}\label{sec:ip-example}
\vspace{-7.5pt}
\input{Tex/2-IP-examples}

\vspace{-5pt}
\section{LTL Property Mining}\label{sec:property-mining}
\vspace{-5pt}
\input{Tex/3-property-mining}

\vspace{-5pt}
\section{Verification of the Implementation}\label{sec:veri}
\vspace{-5pt}
\input{Tex/4-verification}

\vspace{-5pt}
\section{Discussion}\label{sed:discuss}
\vspace{-5pt}
\input{Tex/5-Discussion}

\vspace{-5pt}
\section{Related Work}\label{sec:related}
\vspace{-5pt}
\input{Tex/6-related-work}

\vspace{-7.5pt}
\section{Conclusions}\label{sec:conclu}
\vspace{-5.5pt}
\input{Tex/7-conclusion}

\section*{Acknowledgements}
The authors would like to thank the anonymous reviewers for their constructive comments.
This work was supported in part by the National Natural Science Foundation of China (Nos. 62372304, 62302375, 62192734), and the China Postdoctoral Science Foundation funded project.

\bibliographystyle{unsrt} 
\bibliography{Tex/reference}

\end{document}

%% file: Tex/0-abstract.tex
IP-based software design is a crucial research field that aims to improve efficiency and reliability by reusing complex software components known as intellectual property (IP) components. 
To ensure the reusability of these components, particularly in security-sensitive software systems, it is necessary to analyze the requirements and perform formal verification for each IP component. 
However, converting the requirements of IP components from natural language descriptions to temporal logic and subsequently conducting formal verification demands domain expertise and non-trivial manpower.
This paper presents a case study on software IP components derived from aerospace embedded systems, with the objective of automating the requirement analysis and verification process. 
The study begins by employing Large Language Models to convert unstructured natural language into formal specifications. 
Subsequently, three distinct verification techniques are employed to ascertain whether the source code meets the extracted temporal logic properties. 
By doing so, five real-world IP components from the China Academy of Space Technology (CAST) have been successfully verified.

%% file: Tex/1-introduction.tex
\textbf{Background and Its Importance.} 
Software systems are becoming increasingly complex and diverse, requiring high levels of efficiency and reliability. One way to achieve these goals is to reuse existing software components that have been proven to work well in different contexts. These components are known as intellectual property
components (components of \textit{intellectual property}, or simply \textit{IP components})~\cite{boke2001software,wagner2004strategies,gundabolu2018chip}, and they can range from simple functions to entire subsystems. 
IP-based software design is a crucial research field that aims to facilitate and optimize the reuse of IP components across various software systems~\cite{de2004design,kunkel2003toward,cs2016software}.

However, reusing IP components is not a trivial task. It requires ensuring that the components meet the specific requirements or specifications of the target system, and that they do not introduce any errors or vulnerabilities.
Therefore, it is necessary to analyze the requirements and perform formal verification for each IP component before integrating it into the system.
This process involves converting the natural language descriptions of the requirements into formal logic expressions, and then checking whether the source code of the component satisfies these expressions. 
However, this process is often time-consuming and labor-intensive, as it demands domain expertise and manual intervention.

\input{Figure/overview}

\textbf{Research Problem and Gap.} 
Despite the advancements in IP-based software design, there is still a lack of effective and automated methods for analyzing and verifying IP components, especially in security-sensitive software systems. These systems have strict requirements that must be thoroughly verified to ensure safety and reliability. Typically, these complex requirements are expressed in natural language, incorporating domain knowledge, and require a significant time investment for understanding~\cite{cosler2023nl2spec,sun2023towards,zhai2020c2s}. 
Additionally, these systems often use IP components from different sources and domains, adding to the complexity of verification tasks. 
Therefore, there is a critical need for a comprehensive approach to automated requirements analysis and verification for IP components.

\textbf{Our Objective and Contributions.} 
In this paper, we present a case study on software IP components derived from aerospace embedded systems, with the objective of automating the requirement analysis and verification process. As illustrated in Fig.~\ref{fig:workflow}, our approach consists of two modules: (1) the construction of a prompting engineering framework utilizing Large Language Models (LLMs) to transform unstructured natural language into Linear Temporal Logic (LTL), encapsulating this process into an automated tool to reduce the costs and barriers to formalization application; (2) the use of three distinct verification techniques to ascertain whether the source code meets the extracted temporal logic properties, ensuring the consistency of software requirements and code from various perspectives. By doing so, we successfully verified five real-world IP components from the China Academy of Space Technology (CAST), demonstrating the feasibility and effectiveness of our approach.


%% file: Figure/overview.tex
\begin{figure}[t]
\vspace{-5pt}
\centering
\includegraphics[width=0.8\linewidth]{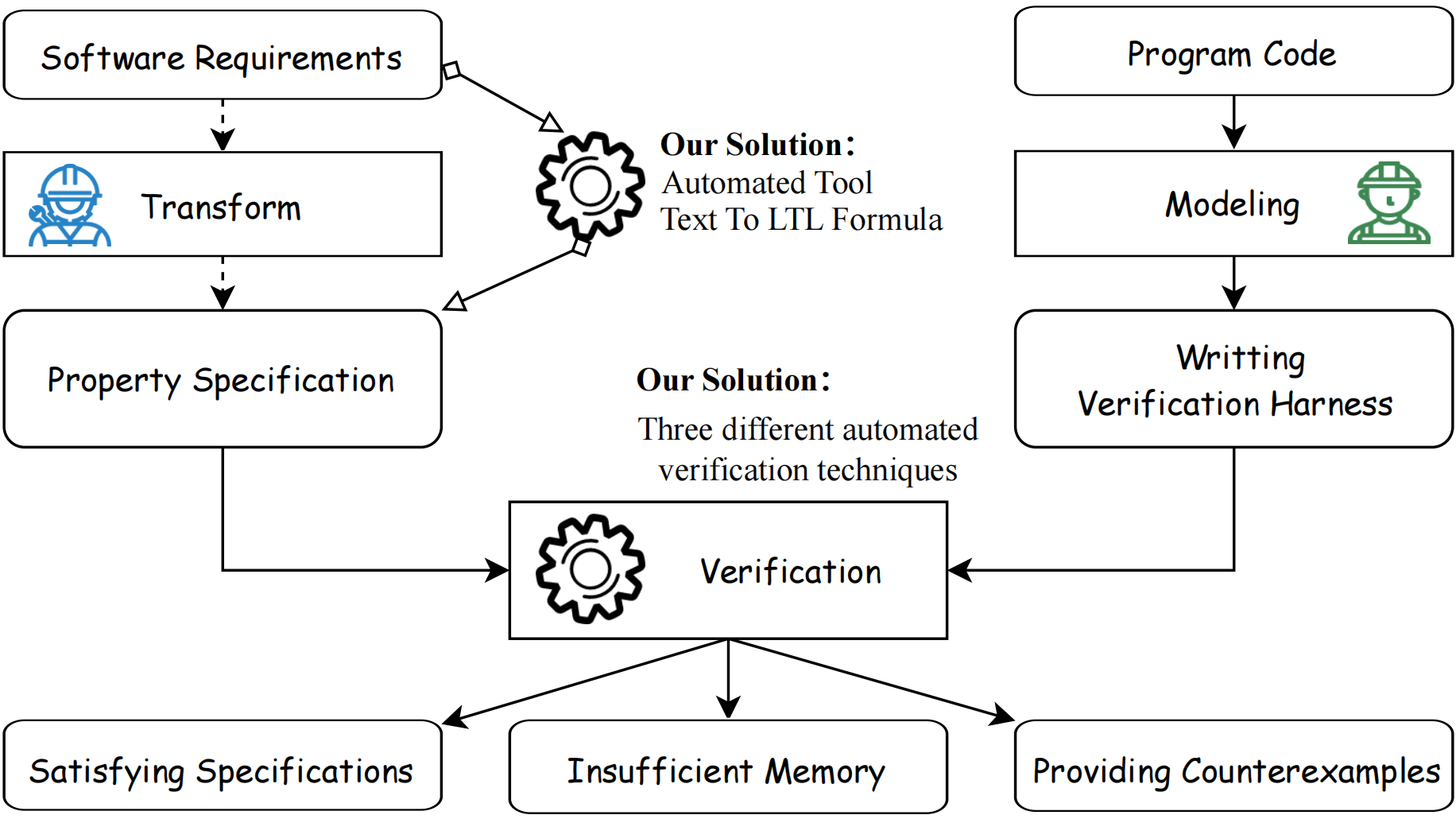}
\setlength{\abovecaptionskip}{5pt}
\setlength{\belowcaptionskip}{-20pt}
\caption{The workflow of our study}
\label{fig:workflow}
\end{figure}

%% file: Tex/2-IP-examples.tex
\smallskip
\textbf{Background of Aerospace Embedded Software IP.}
Functional modules in software can be represented using a \textit{input-output model}. The execution of a functional module involves three steps: input, calculation, and output. Firstly, the input end acquires the necessary computational resources. Then, the computation is performed. Finally, the output end produces the computation result. The interpretation of the input and output may vary, resulting in different software models. In contrast to general software functional modules, software intellectual property (IP)  refers to reusable software knowledge entities that have intellectual property rights. It includes software that has specific functions and is characterized by clear contextual dependencies, interface definitions, comprehensive design documentation, and rigorous quality verification~\cite{crnkovic2010classification}.

A software IP with a proper structure should reflect its key characteristics accurately and comprehensively throughout its creation and utilization. 
It abstracts and consolidates the software knowledge of developers, forming the foundational concept for reusing complex software components. 
Generally, the representation model for software IP consists of the following triplets:
\vspace{-4.5pt}
\begin{equation}\label{eq:ip}
IP = (KL, \; FM, \; IMP)
\vspace{-4.5pt}
\end{equation}
Where $KL$ denotes the \textit{knowledge model}, $FM$ denotes the \textit{formal model}, and $IMP$ denotes the \textit{implementation}.

\input{Figure/IPmodel}

A \textit{knowledge model} is a structured representation of the knowledge within software IP, which is the most abstract and information-rich aspect. 
Knowledge models usually use informal natural language and graphical models, while formal models use formal language. 
The implementation mainly involves program code. From an abstract and refined perspective, the knowledge model covers the entire lifecycle of software IP, providing it with the most extensive content. 
The knowledge model helps developers and users understand software IPs from different perspectives. 
It also enables the search, adaptation, and assembly of software IPs. A \textit{formal model} represents the knowledge model formally. 
Unlike knowledge models, formal models allow a deeper understanding of software IPs. 
They bridge the cognitive gap between software IP creators and users and provide security guarantees for intelligent software synthesis. 
Finally, the program code in the \textit{implementation} must match the description in the formal model, as shown in Fig.~\ref{sec:ip-example}. 
Therefore, to ensure the reusability and trustworthiness of the IP component, we need to verify that:
\vspace{-4.5pt}
\begin{equation}\label{eq:lmp}
IMP \; \models \; FM \; \models \; KL
\vspace{-4.5pt}
\end{equation}
that is, we need to verify that the implementation conforms to the formal properties, and that the formal properties reflect the essential knowledge of the software IP. 
In other words, the implementation, the formal specification, and the knowledge representation of the software IP are consistent and compatible with each other. 
This way, we can ensure that the software IP is developed according to the abstract design and meets the desired functionality and quality.

\input{Table/Table-data-info}
\input{Table/Table-requirement_desc}

\smallskip
\textbf{An Industrial IP Example.}
In the following, we provide an illustrative example of industrial software IP. The purpose of this example is to motivate our research and highlight the complexities involved in analyzing and verifying IP components for aerospace embedded systems.
It is important to note that the software IP components we analyzed and verified were obtained from China Academy of Space Technology (CAST). 
Due to the confidentiality of CAST, we are unable to publicly share these IP components. 
Therefore, the requirement description and code implementation of the IP example presented in our paper has been simplified and abstracted.

The provided Software IP component, namely \texttt{Fg333saCheck},  implements the \textit{\texttt{Fg333} fiber optic gyroscope communication protocol} validation algorithm, which is a method to check whether the data received from a fiber optic gyroscope (FOG) is correct and consistent with the expected format. A FOG is a device that measures the angular velocity of a moving object by detecting the interference of light waves in a closed loop of optical fiber.
There are different types of FOGs, such as closed-loop and open-loop, that use different control and demodulation algorithms. The validation algorithm depends on the specific FOG model and protocol. One example of a FOG protocol is the \texttt{Fg333} protocol, which uses a 19-byte data frame with a frame header, a frame count, a checksum, and other information. The validation algorithm for this protocol involves verifying the data length, the frame count, the frame header, and the checksum of each data frame.
The validation algorithm is important for ensuring the accuracy and reliability of the FOG data, which can be used for various aerospace applications such as navigation, positioning, and attitude control~\cite{lefevre2022fiber}.

The software IP is associated with a comprehensive \textbf{\textit{knowledge model}} that combines fundamental information, domain expert knowledge, and software development process knowledge. This includes the documentation of software requirements and design, which is typically expressed in natural language. In this specific case, the requirements and design documentation are contained within a three-page Word document. We provide below a partial excerpt of the natural language description of the knowledge model, comprising two modules. The first module consists of interface description information for inputs and outputs, as illustrated in Table~\ref{tab:data_info}. This module includes interface function names, types, and explanations. The second module consists of the IP's requirement specification document, as depicted in Table~\ref{tab:requirement_desc}, wherein functional requirements of the IP are documented in natural language.
We can see that the knowledge model is articulated entirely in natural language.
In order to ensure consistency between the code implementation, the formal model, and this knowledge, domain knowledge validators often need to manually convert the information into LTL formulas~\cite{brunello2019synthesis,maidi2000common}. This task requires significant expertise and time, as it necessitates a deep understanding of the knowledge model.

\input{Figure/implementation-model}

The \textbf{\textit{formal model}} of this IP is a specification described using formal language, which has the characteristics of normativity and non-ambiguity. 
Using formal language to describe software IP can minimize the cognitive gap between software IP creators and users, which is beneficial for correctly understanding the meaning of software IP and providing guidance and security guarantees for its reuse.
In general, for a given program $M$, based on Floyd-Hoare logic~\cite{pratt1976semantical,ying2012floyd}, its formal model can be represented as a Hoare triplet: 
\vspace{-4.5pt}
\begin{equation}\label{eq:hl}
\{p\}\; M\; \{q\}
\vspace{-4.5pt}
\end{equation}
Here, $p$ denotes the property that $M$'s input satisfies (precondition), and $q$ represents the property that the output satisfies after the execution of $M$ (postcondition). The assertion $(p, q)$ defines the program $M$'s contract. 
The formal model serves as a bridge between requirements expressed in natural language formats and the implementation of code written in programming languages. It requires the manual abstraction of knowledge models and code implementations and must undergo rigorous formal verification before it can be used.

The \textbf{\textit{implementation}} of a software IP comprises two parts: the interface and the entity, as illustrated in Fig.~\ref{fig:imple-model}. 
The interface represents the visible components of a software IP. Each software IP exposes only one interface to the external world, consisting of input and output ports.
While the entity consists of two parts: private variables and member functions. Private variables are exclusively used internally within the software IP, shared among internal member functions, and remain invisible externally. External access to private variables can only occur through the ports.

\input{Listing/Listing2}

A partial implementation sample of \texttt{Fg333} protocol validationis presented in Listing~\ref{listing:case-study-entity}.
The interface can be declared within the \texttt{.h} file of the software IP code, allowing external environments to access the software IP interface solely through the \texttt{Fg333saCheckFun} function, without requiring knowledge of the interface's implementation details.
The entity of a software IP exists in the form of a \texttt{.c} (source file) or \texttt{.a} (compiled static link library), representing the specific implementation of the software IP's functionality.

%% file: Figure/IPmodel.tex
\begin{figure}[t]
\vspace{-5pt}
\centering
\includegraphics[width=0.65\linewidth]{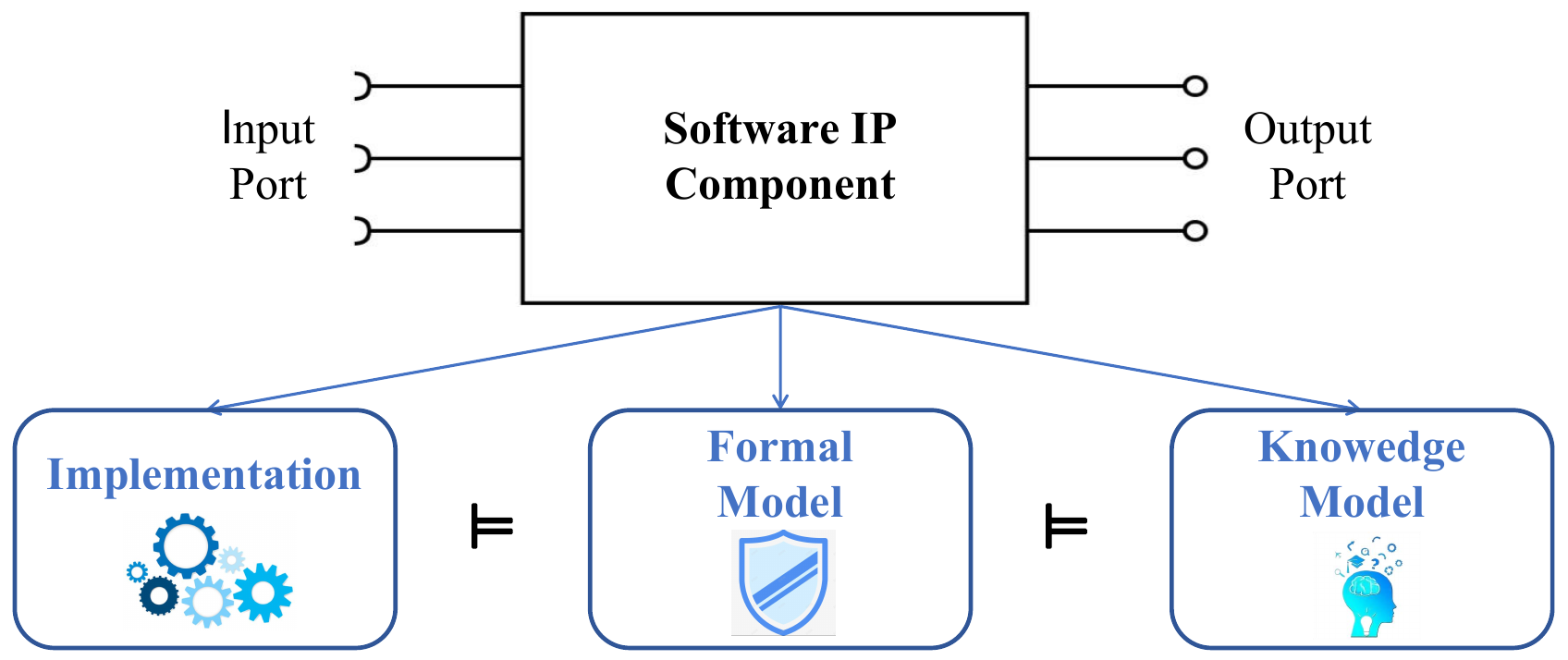}
\setlength{\abovecaptionskip}{5pt}
\setlength{\belowcaptionskip}{-20pt}
\caption{A General Model for Aerospace Embedded Software IP}
\label{fig:ip-model}
\end{figure}

%% file: Table/Table-data-info.tex
\begin{table}[t]
    \vspace{-5pt}
    \centering
    \footnotesize
    \setstretch{0.85}
    \setlength{\tabcolsep}{1pt}
    \captionsetup{justification=centering}
    \setlength{\abovecaptionskip}{2.5pt}
    \setlength{\belowcaptionskip}{-2.5pt}
\caption{\small{Variables in \texttt{struct Fg333saCheck} are presented in a table in the document\iffalse Whether to archive a table or write it directly in a document depends on several factors?\fi}}
\label{tab:data_info}
\begin{tabular}{|l|c|c|l|}
\hline
\textbf{Data\;Name} & \textbf{Data\;Type} & \textbf{Category} & \textbf{Explanation} \\ \hline
\texttt{buffer} & uint8* & input port & The input string buffer \\ \hline
\texttt{rdLen} & uint32 & input port & Length of the input string buffer \\ \hline
\texttt{frm} & uint32 & state\;variable & Frame count \\ \hline
... & ... & state\;variable & ... \\ \hline
\texttt{bComSuc} & uint32 & output port & A flag indicates whether the validation success \\ \hline
\texttt{cntLenRd} & int32 & output port & Continuous read length error count \\ \hline
\texttt{cntHead} & int32 & output port & Continuous frame header error count \\ \hline
\texttt{cntCheck} & int32 & output port & Continuous check error count \\ \hline
\texttt{cntUpdata} & int32 & output port & Continuous data update error count \\ \hline
\texttt{totalLenRd} & int32 & output port & Total read length error count \\ \hline
\texttt{totalHead} & int32  & output port & Total frame header error count \\ \hline
\texttt{totalCheck} & int32 & output port & Total check error count \\ \hline
\texttt{totalUpdata} & int32 & output port & Total data update error count \\ \hline
... & ... & output port & ... \\ \hline
\end{tabular}
\end{table}

%% file: Table/Table-requirement_desc.tex
\begin{table}[t]
    \centering
    \footnotesize
    \setstretch{0.90}
    \setlength{\tabcolsep}{0pt}
    \captionsetup{justification=centering}
    \setlength{\abovecaptionskip}{-5pt}
    \setlength{\belowcaptionskip}{-2.5pt}
\caption{Requirement in Natural Language Description}
\label{tab:requirement_desc}
\begin{quote}
\color{teal}
This IP is used to verify whether the data received by the sensor meets the standard of the \texttt{Fg333} type fiber optic gyroscope communication protocol. The verification rules are as follows:
\vspace{-5pt}
\begin{enumerate}
\item Validate the data length for correctness: if the length is not 19, increment \texttt{cntLenRd} and \texttt{totalLenRd} + 1, and return a validation failure.
\item Validate whether the frame count is updated: if the frame count is not updated, increment \texttt{cntUpdata} and \texttt{totalUpdata} + 1, and return a validation failure.
\item Validate whether the frame header is correct: if the frame header is not \texttt{0xAC12}, increment \texttt{cntHead} and \texttt{totalHead} + 1, and return a validation failure.
\item Validate whether the checksum is correct: if the checksum is correct, return validation success; otherwise, return validation failure.
\end{enumerate}
\end{quote}
\vspace{-20pt}
\end{table}

%% file: Figure/implementation-model.tex
\begin{figure}[t]
\vspace{-5pt}
\centering
\includegraphics[width=0.85\linewidth]{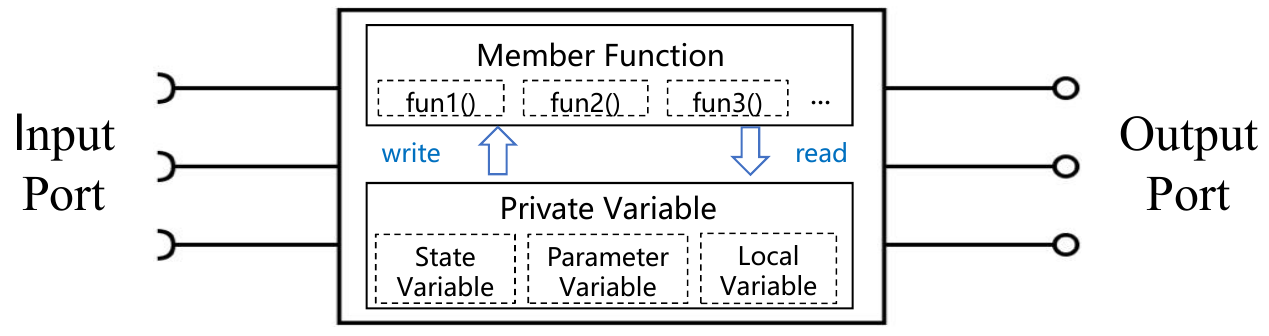}
\setlength{\abovecaptionskip}{5pt}
\setlength{\belowcaptionskip}{-20pt}
\caption{Implementation of Software IP}
\label{fig:imple-model}
\end{figure}

%% file: Listing/Listing2.tex
\definecolor{stubbg}{HTML}{FCF3D5}
\begin{lstlisting}[
    language=c,
    %basicstyle=\linespread{1.0}\tiny,
    morekeywords={var},
    aboveskip=-5pt,
    belowskip=-20pt,
    float=tb,
    caption=Partial Implementation Sample,
    label={listing:case-study-entity},
    escapechar=|,
    numbers=left
]
// Interface.h
extern void Fg333saCheckFun(void *p);

// Entity.c
void Fg333saCheckFun(Fg333saCheck *is) {
	uint08 chksum;
	is->bComSuc = FALSE;
	if (is->rdLen == 19) {
		is->cntLenRd = 0;
		if (is->buffer[17] != (is->frm)) {
			is->cntUpdata = 0;
			is->frm = is->buffer[17];
			if ((is->buffer[0] == 0xAC) && (is->buffer[1] == 0x12)) {
				is->cntHead = 0;
				chksum = CheckSumAdd08(&is->buffer[0], 18);
				if (chksum == is->buffer[18]) {
					is->cntCheck = 0;
					is->bComSuc = TRUE;
				} else {
					is->cntCheck++;
					is->totalCheck++;
				}
			} else {
				is->cntHead++;
				is->totalHead++;
			}
		} else {
			is->cntUpdata++;
			is->totalUpdata++;
		}
	} else {
		is->cntLenRd++;
		is->totalLenRd++;
	}
	return;
}
\end{lstlisting}

%% file: Tex/3-property-mining.tex
In the field of software engineering, the translation of software requirement documents from natural language to formal language is a crucial and complex process. This step is vital for ensuring the accuracy and consistency of requirements, serving as an input for the formal verification of software properties. By utilizing Large Language Models (LLMs) like OpenAI's GPT series~\cite{wu2023brief,kocon2023chatgpt}, Google's BERT~\cite{ravichandiran2021getting}, and T5, which leverage deep learning technology, significant progress has been made in this domain~\cite{cosler2023nl2spec,wen2024enchanting,wen2024automatically,xie2023impact}. These sophisticated tools are specifically designed to comprehend and generate human language. Through the analysis and processing of vast amounts of textual data, these models acquire an understanding of the intricate structures and patterns of language. The key benefits of these models include 1. \textit{Advanced Language Comprehension} - their ability to deeply understand the subtle semantics and context within natural language is crucial for accurately interpreting software requirement documents. 2. \textit{Flexible Text Generation} - these models not only understand natural language but can also generate formal language descriptions that adhere to specific grammatical and formatting requirements. 3. \textit{Adaptability and Versatility} - they can be applied to various software engineering tasks and different programming contexts, handling requirement documents with diverse styles and structures. 4. \textit{Continuous Learning and Improvement} - as these models are continuously trained, their ability to comprehend complex requirements and generate precise formalized descriptions is consistently enhanced. 5. \textit{Automation and Efficiency} - the use of LLMs can reduce the cost of verification efforts and make it easier to adopt formal methods.

While Large Language Models (LLMs) offer a promising solution for translating natural language into formal language, increasing automation in the verification process, reducing costs related to formal methods, and lowering adoption barriers, there are challenges. Knowledge models written in natural language often lack standardization and precision, and not all requirement descriptions can be easily converted into formal specifications. These factors significantly impact the accuracy and reliability of outcomes. For instance, in Table~\ref{tab:requirement_desc} of the knowledge model, the requirement descriptions for the first three functionalities are non-standard and mix natural language with mathematical symbols.

\textbf{Mining Process.} To enhance the accuracy and success rate of transforming natural language into formal language, we have divided the mining of the IP knowledge model into the following three steps:
\vspace{-5pt}
\begin{enumerate}
\item Filter requirement descriptions that contain temporal properties in preparation for translation into Linear Temporal Logic (LTL);
\item Standardize the natural language obtained in the first step, and refine abstract variable descriptive nouns into variable names at the code level;
\item Translate the intermediate results obtained in the $2^{nd}$ step into LTL formulas.
\end{enumerate}
\vspace{-15pt}

\subsection{Step I: Requirements Description Filtering.}
This initial phase involves identifying and selecting requirement descriptions that contain temporal aspects, which are crucial for transformation into LTL. The importance of this step lies in its ability to filter out irrelevant or inapplicable requirements, focusing only on those that can feasibly be transformed into formal specifications. This process includes a detailed examination of the language used in the requirements to determine the presence of temporal elements, ensuring that only suitable requirements proceed to the next phase. 

First, collect a large number of software requirement documents, which should include various types of requirement descriptions, temporal and non-temporal. This data will serve to train an LLM to identify and differentiate between different types of requirement descriptions. In the training data, mark the descriptions that have temporal properties. This marking might involve identifying specific temporal vocabulary (such as `after', `before', `during', \textit{etc}.) or statements with time constraints. Subsequently, use these annotated data to train the LLM. During the training process, the model learns how to distinguish between temporal and non-temporal requirement descriptions based on textual features (vocabulary, grammatical structure, \textit{etc}.). Finally, optimize and validate the model's performance through cross-validation and testing sets to ensure the model accurately identifies requirements descriptions that contain temporal properties.

By the way, our decision to convert natural language into Linear Temporal Logic (LTL) stems from two main reasons. 
Firstly, several advanced formal verification tools rely on LTL specifications for verifying properties, making LTL a suitable input for these tools. 
Secondly, LTL can be easily translated into other temporal logic constraint languages like CTL~\cite{maidi2000common,zhang2021temporal} and PPTL~\cite{ning2022pptl} to provide a more comprehensive depiction of software systems.

\vspace{-5pt}
\subsection{Step II: Natural Language Standardization and Refinement}
Upon isolating the pertinent requirements, the next step is to standardize the natural language used in these requirements. This standardization involves restructuring sentences for clarity, removing ambiguities, and ensuring consistency in terminology. Additionally, this step involves refining abstract variable descriptors into specific, code-level variable names. This refinement is critical for maintaining consistency and ensuring that the translated requirements accurately reflect the intended functionalities in the code.

\input{Figure/Prompt1}

\input{Figure/Prompt2}

Taking the first requirement, Requirement A, from Table~\ref{tab:requirement_desc} as an example: ``\textit{Validate the data length for correctness: if the length is not 19, increment cntLenRd and totalLenRd + 1, and return a validation failure.}" This requirement describes a function that IP Fag333saCheck should possess, merging natural language with mathematical formulas. The term 'length' here abstracts a concept from the code level. We categorize such descriptions as implicit requirement descriptions, which are expressions that are not standardized. Translating them directly into an LTL expression might conceal critical information at the code level and disrupt the subsequent consistency validation. Hence, we have developed a specific prompt engineering approach for the LLM, as illustrated in Fig.~\ref{fig:prompt1}, to transform implicit requirement descriptions into explicit ones. More precisely, program variables mentioned in the requirements should utilize data interfaces as defined in the knowledge model. The contents of Table 1 should be used as a query dictionary to match program variables to their corresponding `Data Name' in the `Explanation' column. The blending of natural language with mathematical formulas should be avoided. Also, function return values should be denoted by TRUE and FALSE for success and failure, respectively. Following this procedure, we obtain the explicit requirement description for Requirement A: ``\textit{Validate the Correctness of Data Length: If the data length reLen is not equal to 19, the values of the variables cntLenRd and totalLenRd will increase by 1 each, and the return value will be FALSE.}" Similarly, we applied the LLM to rewrite the second requirement, Requirement B, from Table 2, resulting in: ``\textit{Validate Frame Count Update: If the frame count \texttt{pbuff} equals \texttt{frm}, then the values of the variables cntUpdata and totalupdata will increase by 1 each, and the return value will be FALSE.}"

\vspace{-5pt}
\subsection{Step III: Translation into LTL Formulas}
The final step in the process involves translating standardized and refined requirements into LTL formulas. This translation requires a thorough understanding of both natural language semantics and formal logic syntax. The goal is to create precise and unambiguous LTL representations of the requirements that are logically sound and true to the original natural language descriptions. This step is vital as it connects informal requirement descriptions with formal specifications, facilitating a more effective application of automated verification processes.

In a similar manner, we construct a prompt engineering framework, utilizing LLMs to translate explicit requirements obtained from the second step into LTL formulas. Fig.~\ref{fig:prompt2} illustrates our specific approach. The process begins with the identification of atomic propositions, translating each significant element or condition in natural language sentences into atomic propositions. An atomic proposition is a basic, indivisible statement in LTL. Subsequently, the translation involves mapping tenses and logical structures from the natural language to corresponding LTL operators, including specific temporal constraints mentioned in the natural language. Ultimately, the LTL formula is constructed by combining atomic propositions with LTL operators, ensuring a rigorous and academically sound translation from natural language to formal logic representation. Taking the first requirement `\textit{Validate the Correctness of Data Length}' as the translation subject, the resultant LTL formula is:
\vspace{0pt}
\begin{equation}\label{formula:firstreq}
\begin{split}
G(\text{reLen} != 19 \rightarrow F(\text{cntLenRd}' = \text{cntLenRd} + 1 \, \\ 
\&\& \; \text{totalLenRd}' = \text{totalLenRd} + 1 \, \\
\&\& \; \text{reVal} = \text{FALSE}))
\end{split}
\vspace{-2.5pt}
\end{equation}

%% file: Figure/Prompt1.tex
\newtcbtheorem[auto counter, number within = section]{cmt}{}{
	colbacktitle = black!60!white, colframe = black!60!white,
	colback = black!5!white,
	fonttitle=\bfseries,
}{t}

\begin{figure}[t]
\centering
\scriptsize
\vspace{-2.5pt}
\begin{cmt*}{Prompt}
\vspace{-5pt}
\textbf{Prompt file}: Transforming non-standard natural language software requirements into standardized natural language descriptions that align with the style of Linear Temporal Logic (LTL) formulas.

\vspace{2.5pt}
\textbf{Objective}: To convert non-standard, vague, or ambiguous software requirements into more coherent, precise, and standardized representations that adhere to the principles of Linear Temporal Logic (LTL).

\vspace{2.5pt}
\textbf{Method}:
\vspace{-5pt}
\begin{enumerate}
\item Understanding Original Requirements: initially comprehend the given non-standard natural language descriptions of software requirements.
\item Identifying Key Elements: Extract key actions, objects, conditions, etc.
\item Altering Mixed Expression Styles: When encountering natural language mixed with mathematical formulas, rewrite the mathematical formulas in a natural language style.
\item Standardizing Language Representations: Reconstruct the original requirements using clear and direct language to express these elements.
\item Applying Temporal Logic Vocabulary: Use LTL concepts like "always", "until", "eventually", etc., to make the representations more aligned with the LTL style.
\item Precisely Describe Variables involved in requirements: Consult the knowledge model tables to transform abstract descriptions into more concrete ones.
\item Providing Conversion Examples: For each requirement, offer examples comparing before and after the transformation.
\item Explaining Conversion Rationale: Briefly explain why such transformations make the requirements more aligned with the style of LTL formulas.
\end{enumerate}
\vspace{-5pt}

\textbf{Example}:
\vspace{-7.5pt}
\begin{itemize}
\item Original Requirement: "When a user clicks a button, the system should display data, but only if the network connection is normal.
\item Standardized Requirement: "The system always displays data when the user clicks a button and the network connection is normal.
\item Explanation: The original requirement uses conditional and temporal language; the transformed requirement more explicitly employs "always" and conditional conjunction, aligning with the expression style of LTL.
\end{itemize}
\vspace{-5pt}

\textbf{Conclusion}: Ensure that all transformed requirements maintain the integrity and accuracy of the original intent while being more suited for LT-style logical analysis.   
\end{cmt*}
\setlength{\abovecaptionskip}{0pt}
\setlength{\belowcaptionskip}{-10pt}
\caption{Prompt:\;Conversion\;of\;Implicit\;Requirement\;Descriptions\;into\;Explicit\;Forms}
\label{fig:prompt1}
\end{figure}

%% file: Figure/Prompt2.tex
\begin{figure}[t]
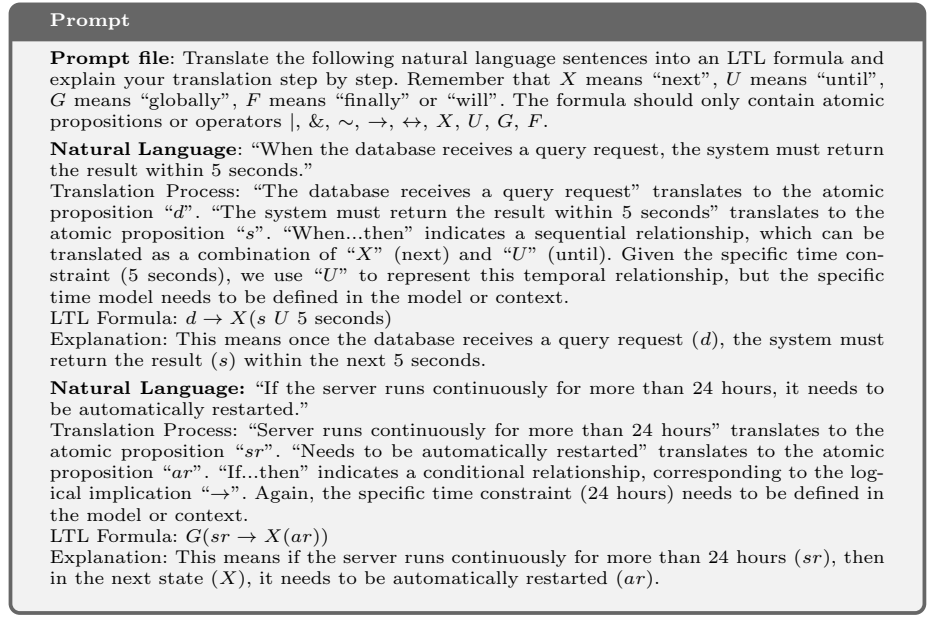

\centering
\scriptsize
\vspace{-2.5pt}
\begin{cmt*}{Prompt}
\vspace{-5pt}
\textbf{Prompt file}: Translate the following natural language sentences into an LTL formula and explain your translation step by step. Remember that \( X \) means “next”, \( U \) means “until”, \( G \) means “globally”, \( F \) means “finally” or “will”. The formula should only contain atomic propositions or operators \( \mid \), \( \& \), \( \sim \), \( \rightarrow \), \( \leftrightarrow \), \( X \), \( U \), \( G \), \( F \).

\vspace{2.5pt}
\textbf{Natural Language}:  “When the database receives a query request, the system must return the result within 5 seconds.”

Translation Process:
   “The database receives a query request” translates to the atomic proposition “$d$”.
   “The system must return the result within 5 seconds” translates to the atomic proposition “$s$”.
   “When...then” indicates a sequential relationship, which can be translated as a combination of “$X$” (next) and “$U$” (until).
   Given the specific time constraint (5 seconds), we use “$U$” to represent this temporal relationship, but the specific time model needs to be defined in the model or context.

LTL Formula: \( d \rightarrow X(s \ U \text{ 5 seconds}) \)

Explanation: This means once the database receives a query request (\( d \)), the system must return the result (\( s \)) within the next 5 seconds.

\vspace{2.5pt}
\textbf{Natural Language:} “If the server runs continuously for more than 24 hours, it needs to be automatically restarted.”

Translation Process:
   “Server runs continuously for more than 24 hours” translates to the atomic proposition “$sr$”.
   “Needs to be automatically restarted” translates to the atomic proposition “$ar$”.
   “If...then” indicates a conditional relationship, corresponding to the logical implication “$\rightarrow$”.
   Again, the specific time constraint (24 hours) needs to be defined in the model or context.

LTL Formula: \( G(sr \rightarrow X(ar)) \)

Explanation: This means if the server runs continuously for more than 24 hours ($sr$), then in the next state ($X$), it needs to be automatically restarted ($ar$).
 
\end{cmt*}
\setlength{\abovecaptionskip}{0pt}
\setlength{\belowcaptionskip}{-10pt}
\caption{Prompt: Translate Explicit Requirements into LTL Formulas}
\label{fig:prompt2}
\end{figure}

%% file: Tex/4-verification.tex
\textbf{Tools used.} Software formal verification tools can be helpful in proving the memory safety and correctness of programs. CBMC~\cite{kroening2014cbmc} is a Bound Model Checker tool used for analyzing programs written in the ANSI-C language.
The bounded model checking technique is a type of software model checking that makes symbolic execution based on SAT procedures.
The bounded model checking examines counterexamples of a specific length and generates a propositional formula that is satisfiable if and only if such a counterexample exists~\cite{biere1999symbolic}.

CPAChecker~\cite{beyer2011cpachecker,lowe2014cpachecker} is a versatile framework that enables users to tailor program analysis to their specific requirements. It allows for the customization of verification methods and configurations based on the code's structural aspects, including loops, recursion, and properties like memory safety and reachability. Verification inputs are created by combining the code and specifications, which are then used to construct program and property models, along with algorithm instances. By utilizing composite CPA analysis methods, CPAChecker explores the reachable space and generates detailed verification reports.

TRACE, our independently developed dynamic verification tool, employs a trace-based runtime verification technique to analyze system behavior by examining its execution traces~\cite{reger2016trace,bubel2023trace}. The tool utilizes the symbol execution tool KLEE~\cite{cadar2008klee,cadar2021klee} to generate test cases covering various code branches and dynamically executes the code, while using the debugging tool GDB to obtain actual execution traces. An execution trace is a sequence of events or states representing the system's activity over time. These traces are compared against a specification of desired or undesired properties, which can be expressed using formalisms like temporal logics and automata. Trace-based verification differs from traditional formal verification techniques such as model checking or theorem proving as it focuses on a subset of possible behaviors and interacts directly with the actual system rather than a formal model. While this approach enhances the scalability and practicality of trace-based verification, it sacrifices exhaustiveness and rigor.

\smallskip
\textbf{Verification Process.} Our verification process for a Software IP component includes the following three steps:
\vspace{-1ex}
\begin{enumerate}
\item Identifying properties to be verified from the requirements;
\item Generating a verification harness that can be compiled, the implementation of the IP component is called within a \texttt{main} function, and the source code of the harness is annotated using the specification language;
\item Executing various verification tools to verify safety properties and functional correctness properties derived from the requirements.
\end{enumerate}
\vspace{-10pt}

\subsection{Step I: Identifying Properties to be Verified}

The activity of selecting properties for verification lacks a well-established systematic approach. However, two primary categories of properties can be subject to verification: \textbf{\textit{safety}} and \textbf{\textit{functional correctness}}, which are expressed through preconditions and postconditions or temporal logic formulas.

\input{Table/Table-safety}

Safety properties encompass areas such as memory safety, integer safety, and numeric overflow. By proactively ensuring that these issues do not occur during program execution, we can ensure the safety and reliability of software systems. To verify these safety properties, assertions can be automatically inserted into the program using tools like CBMC, CPAChecker, and KLEE. 
Table~\ref{tab:safety_property} showcases the safety properties supported by different tools, allowing for the application of bounded model checking, static analysis, and symbol execution.

The derivation of functional correctness properties from software requirements, as elaborated in Section~\ref{sec:property-mining}, involves the application of Hoare Logic notation, as introduced by Hoare~\cite{hoare1969axiomatic}.
This notation necessitates the identification of preconditions and postconditions for each function to express behavior properties accurately.
Temporal logic is employed to articulate temporal properties requiring verification, where one property must be satisfied before verifying another. 
For example, the initial requirement can be represented using an LTL formula, as demonstrated in Formula~\ref{formula:firstreq}.

\vspace{-7.5pt}
\subsection{Step II: Generating Verification Harnesses}
\vspace{-2.5pt}
Unlike most program analysis and verification approaches, software IP is an independent component that requires calling and execution, similar to an API. 
Therefore, a verification harness (also known as a test harness)~\cite{gong2012fast,rocha2008method,beyer2017software} is necessary to drive the verification process.

The verification harness includes the translation of target functional correctness properties into a specific notation that can be executed by each tool.
The concept of contracts is employed, During the process of translating properties to annotations.
The syntax of the annotation depends on the tool being used. In our verification harness, we utilize a \texttt{\_\_ASSUME} and \texttt{\_\_ASSERT} notation. 
The use of these annotations simplifies the process, as the assume and assert clauses are commonly employed in verification tools for setting pre/post-conditions. 
Therefore, it allows us to easily adapt this harness to three different verification tools using macro conversions, as shown at the top of Listing~\ref{listing:veri_harness} (Lines 1-15). 
For instance, CBMC utilizes Cprover annotations to define the properties that a program should satisfy. 
Cprover annotations consist of keywords and pragmas that are added as comments in the source code. 
An example of an annotation is \texttt{\_\_CPROVER\_assert(condition, message)}, which verifies whether a condition holds at a specific program point.
Our \texttt{\_\_ASSERT} macro can be converted to the Cprover annotation, as shown in Line 3 of Listing~\ref{listing:veri_harness}. 
Similarly, when using KLEE to generate test input for the runtime verification tool TRACE, there is a need to simplify the most commonly used intrinsic, \texttt{klee\_make\_symbolic}, which creates an unconstrained symbolic object.
This principle of macro usage also applies to KLEE and CPAChecker.

Listing~\ref{listing:veri_harness} provides an example of a verification harness that targets the verification of the first requirement mentioned in Table~\ref{tab:requirement_desc} of Section~\ref{sec:ip-example}. 
This verification harness is partly generated based on templates and follows a standard structure.
Firstly, it declares and initializes the necessary variables that the IP component uses (Lines 18-27), which are similar to the variables that need to be made symbolic in KLEE.
Then, it includes the preconditions for the target properties (Line 30).
Next, it calls the IP component to be verified (Line 33), similar to using the API interface. 
Lastly, the expected postconditions are specified.

\input{Listing/Listing3}

When an IP component involves multiple properties for verification, they can be integrated within the same verification harness if they share the same preconditions.
However, if the properties have distinct preconditions, a new verification harness is necessary for each unique precondition.
Subsequently, it is crucial to examine the union of the preconditions from all verification harnesses to ensure comprehensive coverage of all possible preconditions.
For safety properties, a verification harness without preconditions and postconditions suffices. 
To verify safety properties, we have developed a tool that automatically generates such verification harnesses using lightweight static analysis and pattern matching. 
Although our properties mining process contributes partially to filling in the preconditions and postconditions, we still need to manually input the functional correctness properties into the verification harnesses, due to the significant engineering effort involved in our underlying implementation.

\vspace{-5pt}
\subsection{Step III: Executing three Verification Tools}

Once all verification harnesses have been prepared, we initially execute the verification harness without any preconditions or postconditions to verify safety properties.
Table~\ref{tab:safety_result} presents a detailed description of the results obtained from proving safety properties in five software IPs, utilizing CBMC, CPAChecker, and KLEE.
These five software IPs each have a verification harness but do not include any pre/post-conditions. Each IP consists of approximately one hundred lines of code, which encompasses arrays, pointers, floating-point operations, \textit{etc}.
Three IPs have been successfully proven to be safe using all three techniques.
Only some cases of integer overflow and floating-point arithmetic overflow have been found in \texttt{Fg333saUnpack} and \textit{Fg333saCheck}.
It is worth noting that both CBMC and CPAChecker reported the floating point arithmetic overflow, while only CBMC reported the integer overflow. 
However, KLEE did not identify these two issues. 
After conducting a manual analysis, we discovered that these two issues were indeed potential defects that occurred only under specific preconditions.
This highlights the importance of using multiple tools to verify safety properties.

\input{Table/Table-safety-result}
\input{Table/Table-functional-result}

Table~\ref{tab:functional_result} presents the verification results for five IPs. 
It shows the number of functional properties that were examined and the corresponding number of properties that were successfully verified by each tool (indicated as \texttt{Proved}/\texttt{All}).
The properties consist of 20 temporal logic properties identified through LLM-based automatic mining and 30 manually added pre/post-conditions. Of the 20 temporal logic properties, TRACE+KLEE successfully verified 17. Similarly, out of the 30 pre/post-condition pairs, only 20 were successfully verified. Manual analysis revealed that unsuccessful verifications were influenced by factors like multiple loop structures, floating point arithmetic, and quantifiers. This underscores the need for enhancements in both verification capability and scalability in the context of industrial scenarios.

%% file: Table/Table-safety.tex
\begin{table}[t]
    \vspace{-2.5pt}
    \centering
    \footnotesize
    \setstretch{0.85}
    \setlength{\tabcolsep}{5pt}
    \captionsetup{justification=centering}
    \setlength{\abovecaptionskip}{2.5pt}
    \setlength{\belowcaptionskip}{0pt}
    \vspace{-5pt}
    \caption{Safety properties that Different Tools can Automatically Verify}
    \label{tab:safety_property}
\begin{tabular}{|l|c|c|c|}
\hline
\textbf{\qquad\qquad Safety Property}                         & \textbf{CBMC} & \textbf{CPAChecker} & \multicolumn{1}{c|}{\textbf{KLEE}} \\
\hline
Out of Bounds Array Access & \ding{52} & \ding{56} & \ding{52} \\ 
\cline{1-4} 
Invalid Pointer Dereference & \ding{52} & \ding{52} & \ding{52} \\
\cline{1-4} 
Division by Zero & \ding{52} & \ding{56} & \ding{52} \\
\cline{1-4} 
Integer Overflow & \ding{52} & \ding{52} & \ding{56} \\ 
\cline{1-4} 
Pointer Arithmetic Overflow & \ding{52} & \ding{56} & \ding{56} \\ 
\cline{1-4} 
Floating Point Arithmetic Overflow & \ding{52} & \ding{56} & \ding{56} \\ 
\cline{1-4} 
Shift Operation Overflow & \ding{52} & \ding{56} & \ding{52} \\ 
\cline{1-4} 
Memory Leak & \ding{52} & \ding{56} & \ding{52} \\ 
\cline{1-4} 
Termination & \ding{56} & \ding{52} & \ding{56} \\ 
\cline{1-4} 
Data Race & \ding{56} & \ding{52} & \ding{56} \\ 
\cline{1-4} 
Dead Lock & \ding{56} & \ding{52} & \ding{56} \\ 
\hline
\end{tabular}
\vspace{-15pt}
\end{table}

%% file: Listing/Listing3.tex
\definecolor{stubbg}{HTML}{FCF3D5}
\begin{lstlisting}[
    language=c,
    %basicstyle=\linespread{1.0}\tiny,
    morekeywords={var},
    aboveskip=-5pt,
    belowskip=-20pt,
    float=tb,
    caption=A Verification Harness to Verify the First Requirement,
    label={listing:veri_harness},
    escapechar=|,
    numbers=left
]
#ifdef __CBMC__
#define __ASSUME(condition) __CPROVER_assume(condition)
#define __ASSERT(condition) __CPROVER_assert(condition, "")
#elif __CPACHECKER__
#define __ASSUME(condition) __VERIFIER_assume(condition)
#define __ASSERT(condition) __VERIFIER_assert(condition)
#elif __KLEE__
#define __ASSUME(condition) klee_assume(condition)
#define __ASSERT(condition) assert(condition)

void __MAKE_SYMBOLIC(void *addr, size_t nbytes, const char *name){
#ifdef __KLEE__
    klee_make_symbolic(addr, nbytes, name);
#endif
}

void main() {
    // Initialize variables
    Fg333saCheck pIp;
    __MAKE_SYMBOLIC(&pIp, sizeof(Fg333saCheck), "pIp");

    unint08 buffer[19];
    __MAKE_SYMBOLIC(buffer, sizeof(buffer), "buffer");
    pIp.buffer = &buffer[0];

    int32 cntLenRd_old = pIp.cntLenRd;
    int32 totalLenRd_old = pIp.totalLenRd;
    
    // Preconditions
    __ASSUME(pIp.rdLen != 19);

    // Call the IP interface to be verified
    Fg333saCheckFun(&pIp);

    // Postconditions
    __ASSERT(pIp.bComSuc == FALSE);
    __ASSERT(pIp.cntLenRd == cntLenRd_old + 1);
    __ASSERT(pIp.totalLenRd == totalLenRd_old + 1);
}
\end{lstlisting}

%% file: Table/Table-safety-result.tex
\begin{table}[t]
\vspace{-5pt}
    \centering
    \footnotesize
    \setstretch{0.9}
    \setlength{\tabcolsep}{1.5pt}
    \captionsetup{justification=centering}
    \setlength{\abovecaptionskip}{2.5pt}
    \setlength{\belowcaptionskip}{-2.5pt}
    \caption{Verification results of 5 IPs in Safety Properties}
    \label{tab:safety_result}
\begin{tabular}{l|c|l|c|c|c}
\hline
\textbf{IP Name}     & \begin{tabular}[c]{@{}c@{}}\textbf{LoC}\\{\scriptsize{(Source+Harness)}}\end{tabular} & \textbf{ \ Code Feature} & \scriptsize\textbf{CBMC} & \scriptsize\textbf{CPAChecker} & \multicolumn{1}{c}{\scriptsize\textbf{KLEE}} \\
\hline
\texttt{Fg333saCheck} & 140+35 & {\scriptsize{pointer, array, loop}} & \ding{56} {\tiny\textbf{(IO)}} & \ding{52} & \ding{52} \\ 
\hline
\texttt{Fg333saUnpack} & 77+35 & {\scriptsize{floating point arithmetic}} & \ding{56} {\tiny\textbf{(IO, FPO)}} & \ding{56} {\tiny\textbf{(FPO)}} & \ding{52} \\
\hline
\texttt{PowerOnJudge} & 85+33 & {\scriptsize{pointer}} & \ding{52} & \ding{52} & \ding{52} \\
\hline
\texttt{GyroChooseFun} & 50+58 & {\scriptsize{pointer, array, loop}}  & \ding{52} & \ding{52} & \ding{52} \\ 
\hline
\texttt{AttiCal} & 68+44 & {\scriptsize{pointer, array}} & \ding{52} & \ding{52} & \ding{52} \\ 
\hline
\end{tabular}
    \begin{tablenotes}
        \scriptsize
        \item[*] \ding{52} indicates that the tool successfully verifies all the supported safety properties, as outlined in Table~\ref{tab:safety_property}. While \ding{56} indicates the tool reports that some safety properties cannot be satisfied. "\textbf{IO}" refers to integer overflow, and "\textbf{FPO}" stands for floating point arithmetic overflow.
	\end{tablenotes}
\vspace{-10pt}
\end{table}

%% file: Table/Table-functional-result.tex
\begin{table}[t]
    \centering
    \footnotesize
    \setstretch{0.85}
    \setlength{\tabcolsep}{10pt}
    \captionsetup{justification=centering}
    \setlength{\abovecaptionskip}{2.5pt}
    \setlength{\belowcaptionskip}{5pt}
    \caption{Verification results of 5 IPs in Functional Correctness Properties}
    \label{tab:functional_result}
\begin{tabular}{l|c|c|c}
\hline
\multirow{2}{*}{\textbf{IP Name}} & \multicolumn{2}{c|}{\begin{tabular}[c]{@{}c@{}}\textbf{Pre/post-conditions}\\{(\texttt{Proved}/\texttt{All})}\end{tabular}} & \begin{tabular}[c]{@{}c@{}}\textbf{Temporal logic properties}\\{(\texttt{Proved}/\texttt{All})}\end{tabular} \\
\cline{2-4}
 & \scriptsize\textbf{\;\;CBMC} & \scriptsize\textbf{CPAChecker} & \multicolumn{1}{c}{\scriptsize\textbf{TRACE+KLEE}} \\
\hline
\texttt{Fg333saCheck} & 5/5 & 3/5 & 8/8 \\ 
\texttt{Fg333saUnpack} & 1/7 & 1/7 & 0/1 \\
\texttt{PowerOnJudge} & 6/6 & 6/6 & 9/9 \\
\texttt{GyroChooseFun} & 4/7 & 4/7 & 0/1 \\ 
\texttt{AttiCal} & 4/5 & 4/5 & 0/1 \\ 
\hline
\qquad All & 20/30 & 18/30 & 17/20 \\
\hline
\end{tabular}
\vspace{-10pt}
\end{table}

%% file: Tex/5-Discussion.tex
Our work has several implications for the field of security-critical IP-based software development, particularly in the domain of aerospace embedded systems.
First, our work shows that Large Language Models can be used to automate the requirement analysis process, reducing the time and effort required to convert natural language descriptions into formal logic expressions. 
Second, our work demonstrates that verification harnesses can be automatically generated for each IP component, simplifying the verification process and ensuring consistency and completeness. 
Third, our work reveals that different verification techniques can be applied to the same IP component, providing complementary perspectives and insights on the correctness and robustness of the component. 
Our work also provides a novel and comprehensive solution that seamlessly integrates requirement analysis and verification.
Note that our work differs from other related studies in several aspects. 
For instance, unlike other work that used predefined templates or rules to convert natural language into formal logic, our work leverages Large Language Models to perform this task in an interactive and flexible way. 
Furthermore, unlike some work that relied on a single verification technique, our work employs three different verification techniques, namely model checking, static analysis, and symbolic execution, to verify the same IP component, enhancing the confidence and reliability of the verification results.

However, our work also has some limitations and challenges that need to be considered and addressed in future research.
One limitation is that our work depends on the quality and availability of the natural language descriptions of the IP components, which may not always be clear, complete, or consistent. 
Therefore, future work could explore ways to improve the natural language processing and understanding capabilities of the Large Language Models, or to incorporate other sources of information, such as diagrams, specifications, or documentation, to enhance the requirement analysis process. 
Another challenge is that our work requires a large amount of computational resources and time to perform the verification process, especially for large and complex IP components. 
Therefore, future work could investigate methods to optimize the verification process, such as parallelization, abstraction, or approximation, to reduce the computational cost and time. 
A third challenge is that our work does not consider the interactions and dependencies among different IP components, which may affect the overall functionality and performance of the system.

%% file: Tex/6-related-work.tex

\noindent\textbf{Property Mining.} 
How to acquire better-automated tool support for efficiently transforming natural language documents into formalized formulas has emerged as a hot research topic in the field of formalization in recent years. Andrew et al. from Princeton University proposed the automated software verification toolchain VST~\cite{Cao2018VST-Floyd,Lin2008Model-based}, which aims to achieve this objective. It requires users to manually specify properties for verification. The process involves defining preconditions, postconditions, and loop invariants to facilitate automated program verification with the aid of VST. In the formalization projects of garbage collection and the Verve~\cite{Yang2011Safe} system conducted at Microsoft, a plethora of tools were employed to support the automated verification of properties. This included type-preserving compilers and automatic theorem provers, significantly enhancing the efficiency of property verification tasks. Beyer and colleagues from Freie Universität Berlin introduced an automatic verification tool based on abstract interpretation, CPAchecker~\cite{beyer2011cpachecker,lowe2014cpachecker}, which necessitates manual insertion of assertions and error labels to represent properties for verification. The correctness of software programs is demonstrated through the reachability analysis of target states. In our research, we have developed a prompting engineering framework that leverages LLMs to translate natural language into LTL formulas. To enhance the accuracy of the translation, we conducted extensive preprocessing of the natural language inputs. This involved filtering out natural language statements that embody temporal properties and converting non-standard natural language descriptions into standardized ones.

\noindent\textbf{Program Verification.} 
Works that have utilized the CBMC, CPAChecker, and runtime verification tools have reported successful applications in various industries. 
For instance, the CBMC has recently demonstrated its effectiveness in verifying memory safety in the boot code of AWS Data Centers~\cite{cook2018model} as well as in Amazon Web Services~\cite{chong2021code}. 
In another study~\cite{kim2008unit}, the CBMC was employed to detect previously undetected bugs in the driver software for flash memory.
The CBMC has also been used in conjunction with field testing~\cite{post2009linking}. 
These findings collectively support the notion that the CBMC tool is developed sufficiently for industrial usage. 
CPAChecker has been applied to Google App Engine Cloud~\cite{beyer2014software}.
Notably, other works in the literature compare static analysis tools using either CBMC or CPAChecker, although not specifically comparing the CBMC and CPAChecker tools~\cite{beyer2014status,rocha2012understanding,collavizza2008comparison,chen2018scalable}.
However, the main difficulties of using CBMC and CPAChecker, as well as their potential complementary usage, are not well-established in the literature.
Additionally, many existing works propose different methods and tools for runtime verification~\cite{omer2023runtime,reger2016trace,convent2018hardware,bubel2023trace}. 
In our study, we employ our proprietary tool, Trace, for the verification of the temporal logic properties in aerospace-embedded software IP. 
This decision stems from the fact that CBMC and CPAChecker have proven to be ineffective in verifying the temporal logic property in practical scenarios. 
We distinguish our work by showcasing the integration of three widely utilized verification techniques.
Through this combination, we achieve the successful verification of safety properties and functional correctness in industrial programs within the aerospace field.

%% file: Tex/7-conclusion.tex
In conclusion, we presented a case study on software IP components derived from aerospace-embedded systems, and proposed a novel and comprehensive solution for the practical requirement analysis and verification process.
We employ Large Language Models to translate natural language requirements into program specifications and utilize three different verification techniques to verify the corresponding implementation.
We have successfully verified five real-world IP components from CAST and demonstrated the feasibility and effectiveness of our solution.
As for future work, we are considering developing techniques to verify the integration and combination of multiple IP components.
We also plan to conduct more experiments on more real-world IP  components from CAST.